\documentclass{epl}
\usepackage{graphicx}

\title{Structural distortions and orbital ordering in 
\chem{LaTiO_3} and \chem{YTiO_3}
}
\shorttitle{Distortions in \chem{LaTiO_3}, \chem{YTiO_3}}
\author{
  S.~Okatov\inst{1}\inst{2}\inst{3} \and
  A.~Poteryaev\inst{1}\inst{4} \and
  A.~Lichtenstein\inst{3}}  

\institute{
  \inst{1} Research Institute for Materials, University of Nijmegen, The Netherlands\\
  \inst{2} Institute of Solid State Chemistry, UrD of RAS, Ekaterinburg, Russia\\
  \inst{3} I. Institute of Theoretical Physics, University of Hamburg,  Germany\\
  \inst{4} Centre de Physique Th\'eoretique, Ecole Polytechnique, Palaiseau, France
}


\begin{document}

\maketitle

\begin{abstract}
Theoretical investigations of the electronic, magnetic and structural
properties of LaTiO$_3$ and YTiO$_3$ have been made. In the framework of GGA and
GGA+U scheme we analyzed the effect of the local Coulomb interaction
($U$) value on the atomic forces acting in the experimental structure.
The optimal parameters of the electron-electron on-site interactions as
well as the orbital configurations and magnetic properties are
determined. 

\end{abstract}

\section{Introduction}

Transition-metal perovskite oxides LaTiO$_3$ and YTiO$_3$ are classical
Mott-Hubbard insulators. In spite of the fact that they are
formally isoelectronic with one 3$d$ electron in the $t_{2g}$
shell LaTiO$_3$ has a G-type antiferromagnetic (AFM) structure whereas
YTiO$_3$ has a ferromagnetic (FM) one. The most unusual feature of
these compounds relates with a nearly isotropic magnon spectrum in both
titanates despite of their distorted crystal structures
\cite{b.keimer,b.ulrich}.

There are two theoretical models aimed to explain these effects in LaTiO$_3$.
One of them originates from the work of Kugel and Khomskii
\cite{b.kugel} and operates in terms of the lattice distortions and an
orbital ordering (OO). Another one is proposed by Khalliulin and Maekawa
\cite{b.khaliullin} and describes the above effects by the model of
orbital liquid (OL) where the fluctuations of orbital degrees of freedom
play an important role. According to \cite{b.khaliullin} the energy of
the orbital fluctuations is estimated to be $W_{\rm orb} \sim 160$~meV.
Thus OL model is applicable only if the splitting of the $t_{2g}$
orbital in the crystal field $\Delta$ is significantly smaller than
$W_{\rm orb}$. Otherwise the nearly isotropic exchange is likely to be
due to an orbital ordering with a peculiar orbital configuration
\cite{b.haverkort}.

The OL model was supported by neutron experiments of Keimer {\it et al}
\cite{b.keimer} where no OO in LaTiO$_3$ was found. Fritsch with co-workers
\cite{b.fritsch} made heat capacity and magnetic measurements. Having
supposed the nearly cubic structure of LaTiO$_3$ they concluded that the
energy of the spin-orbit coupling for Ti-sites was $E_{\rm SO}=30$~meV
and the crystal field effect was small. The theoretical estimations
made by Solovyev in Ref. \cite{b.solovyev2} in the framework of the
local spin density approximation (LSDA) and LSDA+U theory showed that
$\Delta$ in LaTiO$_3$ was as small as 49~meV and this value was of the same
order of magnitude as the energy of the spin-orbit (SO) coupling $E_{\rm
SO}=23$~meV. 

Recent X-ray studies \cite{b.cwik,b.hemberger} revealed a sizable
deformation of the TiO$_6$ octahedra and deviation from the cubic
symmetry in LaTiO$_3$ below $T_{\rm N}$. The splitting of the $t_{2g}$-
levels due to crystal deformations was estimated in \cite{b.cwik} to be
about 240~meV, and 120-300~meV in \cite{b.haverkort}. This was also
supported in the {\it ab-initio} calculations \cite{b.pavarini} in the
framework of LDA and LDA+DMFT theory, where $\Delta$ was reported to be
$\sim 200$~meV. Those values of $\Delta$ were larger than $W_{\rm orb}$
and could provide ordering of the $t_{2g}$ orbitals. The OO-model has
had a support in the {\it ab-initio} theoretical studies
\cite{b.mizokawa,b.sawada,b.mochizuki,b.pavarini} for both YTiO$_3$ and LaTiO$_3$
as well as in the direct observations of the orbital ordering in YTiO$_3$
\cite{b.ichikawa,b.akimitsu,b.kiyama}, and in LaTiO$_3$ \cite{b.kiyama} from
the NMR spectra analysis.

The physical properties of both LaTiO$_3$ and YTiO$_3$ are strongly sensitive to
the chemical composition \cite{b.crandles}. The highest value of the
N\'eel temperature in LaTiO$_3$ was reported by Cwik {\it et al} \cite{b.cwik}
and Hemberger with co-authors \cite{b.hemberger} to be $T_{\rm
N}=146$~K, while the magnetic moment was $\mu=0.57\mu_{\rm B}$
\cite{b.cwik}. This magnetic moment disagreed with the theoretical
estimations of more than $0.8\mu_{\rm B}$
\cite{b.solovyev1,b.solovyev2}. For YTiO$_3$ the Curie temperature amounts
to 30~K and $\mu=0.84\mu_{\rm B}$ \cite{b.hester,b.akimitsu,b.kiyama}.


According to the OO model the difference in the magnetic properties of
LaTiO$_3$ and YTiO$_3$ can be attributed to the different types of distortions
(see for example Refs. \cite{b.cwik,b.solovyev2,b.mochizuki}). The main
distortion in YTiO$_3$ has the Jahn-Teller type, while the GdFeO$_3$-type
distortion dominates in LaTiO$_3$. In this letter we investigate the nature
of those crystal distortions and the effects of OO on electronic and
magnetic properties of  the LaTiO$_3$ and YTiO$_3$ taking the effects of local
Coulomb correlations into account.

\section{Methods and models}

Both LaTiO$_3$ and YTiO$_3$ have a P$bnm$ crystal structure. The lattice
constants and atomic positions are taken from \cite{b.hester,b.cwik}.
The structural parameters for LaTiO$_3$ used in the calculations correspond
to the temperature below $T_{\rm N}$ ($T=8$~K). The examination of the
YTiO$_3$ is made for the high temperature structure ($T=293$~K) only because
there is no structural data for YTiO$_3$ at $T<T_{\rm C}$. The crystal
structure relaxations are not performed but the calculated forces on
atoms for the experimental atomic positions are used for the
verification of the agreement between theory and experiment. The
orthorhombic unit cell used in our calculations consists of four formula
units. This cell allows us to consider both FM and AFM structures with
different types of the magnetic ordering.


For electronic structure calculations we use the projected augmented
wave (PAW) method \cite{b.kresse,b.blochl} in the framework of the
density functional theory (DFT). The exchange-correlation correction is
taken into account within the general gradient approximation (GGA)
\cite{b.perdew}. In order to treat the effect of local Coulomb
interactions in a partially filled $3d$ band of Ti the LDA+U method is
applied \cite{b.dudarev,b.anisimov}. The Brillouin zone (BZ) is sampled
with the $7\times 7\times 5$ mesh with its origin at $\Gamma$ point. The
valence states include $2s$, $2p$ for O, $4p$, $4d$, $5s$ for Y, $3p$,
$3d$, $4s$ for Ti and $5p$, $5d$, $6s$ for the La-atoms. The cutoff
energy of the plane-wave expansion is 400~eV.

\section{Results and discussions}

\begin{figure}
\begin{tabular}{cc}
LaTiO$_3$ -- AFM&YTiO$_3$ -- FM\\
\resizebox{7cm}{!}{\rotatebox{0}{\includegraphics{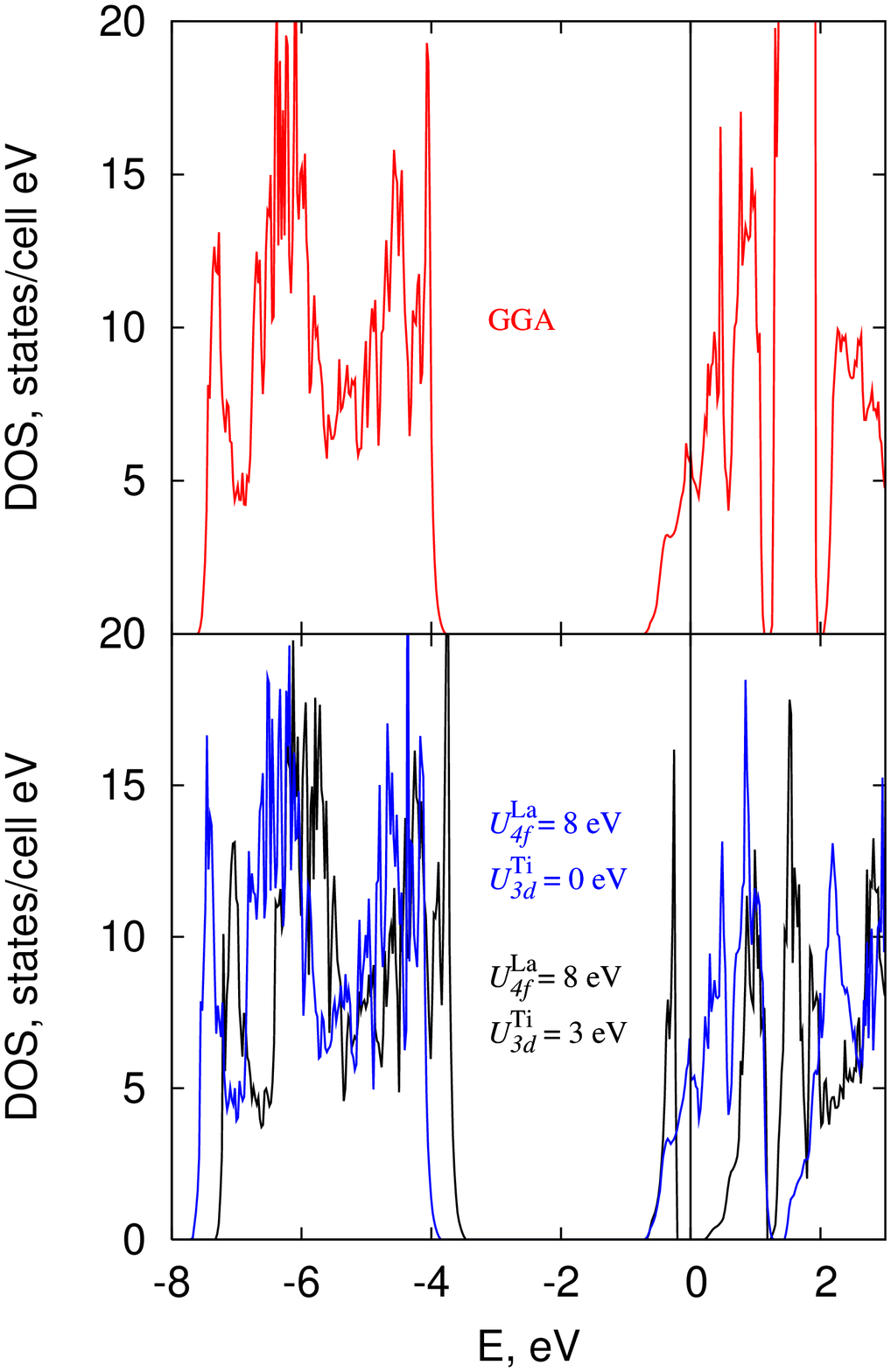}}}&
\resizebox{7cm}{!}{\rotatebox{0}{\includegraphics{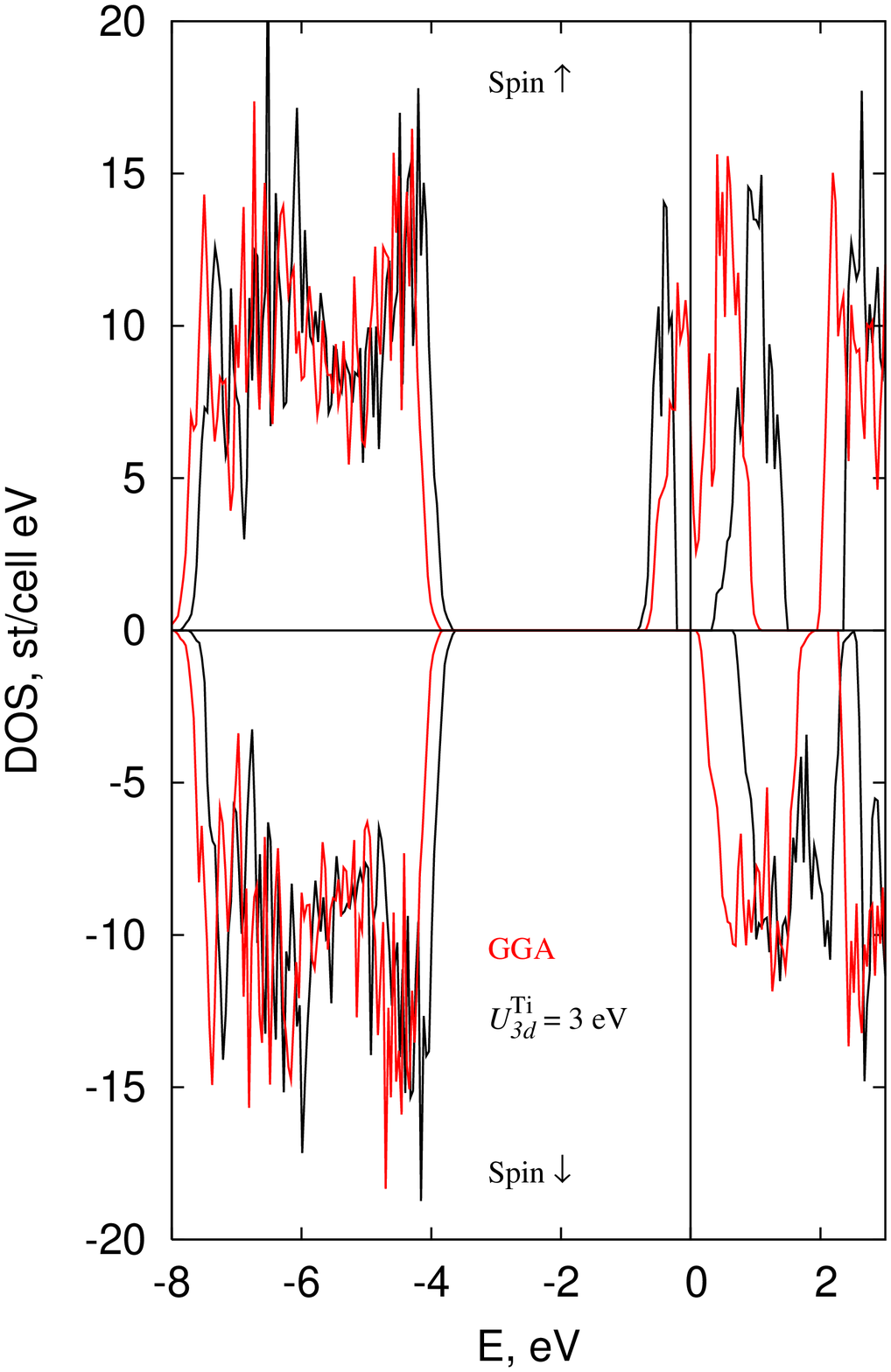}}}\\
\end{tabular}

\caption{
({\it color online}) DOS of the AMF (G-type) LaTiO$_3$ ({\it left}) and FM
YTiO$_3$ ({\it right}) in the framework of the GGA ({\it red}) and GGA+U
({\it blue} and {\it black}, see text) respectively. Upper and lower
panels in YTiO$_3$ are up and down spin states.}


\label{f.dos}
\end{figure}

The results of our GGA calculations for LaTiO$_3$ and YTiO$_3$ are presented in
fig.~\ref{f.dos}. We have found the non-magnetic ground state for LaTiO$_3$
and ferromagnetic ground state for YTiO$_3$ with the value of the magnetic
moment $0.8\mu_{\rm B}$ per Ti atom. This magnitude of the magnetic moment is
very close to the experimental value of $0.85\mu_{\rm B}$
\cite{b.hester,b.akimitsu,b.kiyama}.


In the right part of fig.~\ref{f.dos} the total density of states (DOS)
for ferromagnetic YTiO$_3$ calculated within GGA approach is shown by red
color. It consists of three well separated sets of bands. The lowest in
energy band lies from -8 till -4~eV and is predominantly of O-$2p$
character. The $t_{2g}$ states of Ti are strongly polarized. Spin up
crosses the Fermi level whereas the spin down is completely empty and
lies above the Fermi energy, thus YTiO$_3$ is a half magnetic ferromagnet in
our GGA calculations. The bandwidth of the $t_{2g}$ states is about
1.8~eV for both spin channels. The $e_g$ states of Ti and the rest
unoccupied states (O$3s$, Ti$4p$, etc.) are higher in energy and
separated from $t_{2g}$ states by a small gap. Our results are in good
agreement with the early study of this compound \cite{b.fujitani,
b.sawada}.

In the upper left part of fig.~\ref{f.dos} the total DOS for LaTiO$_3$ is
shown. In this case the DOS is non-spin-polarized and consists of four
parts. As in YTiO$_3$ the fully occupied band of O-$2p$ character is located
from -7.8 to -4~eV. The $t_{2g}$ states cross the Fermi level and have
the bandwidth about 1.9~eV. The $e_g$ states lie starting from 2~eV. In
the energy region between $t_{2g}$ and $e_g$ states the La-$4f$ band is
located. This band ordering Ti-$t_{2g}$, La-$4f$, Ti-$e_g$ is not
compatible with the results of early calculations
\cite{b.solovyev1,b.solovyev2,b.pavarini,b.nekrasov,b.mizokawa} and with
the experimental data \cite{b.sarma}. The simplest way to fix this
problem is to apply GGA+U approach for the La-$4f$ states. The results
of the calculation of LaTiO$_3$ with a value of Coulomb repulsion $U^{\rm
La}_{4f}=8$~eV are shown on the left down part of fig.~\ref{f.dos} by
blue color. The value of Coulomb interaction used is in good agreement
with the experimental and theoretical estimations
\cite{b.solovyev1,b.okimoto,b.mizokawa}. The obtained band ordering is
now in consistency with the previous studies \cite{b.pavarini}. In the
rest part of the paper we will always use this result as a starting
point for the following considerations without explicit mentioning and
will refer to this calculation as GGA.

According to experiments \cite{b.okimoto,b.sarma} both LaTiO$_3$ and YTiO$_3$ are
classical Mott-insulators with the band-gaps ($\Delta E_{\rm g}$) of 0.2
and 1.2~eV respectively whereas in our GGA calculations they are metals.
We have found also that the forces acting on the oxygen atoms in the GGA
are quite big ($\sim$ 0.5~eV/\AA) for the experimental crystal
structure.

It is well known that in the systems where the value of kinetic energy
is comparable with the value of one site Coulomb interaction and strong
electron-electron correlations take place, GGA (or LDA) approximation
may fail in the prediction of the correct ground state. The reason is an
underestimation of the strong correlations in DFT. In order to take into
account the one site Coulomb interaction we use the LDA+U approach which
has orbitally $lm$-dependent potential and describes well the insulating
compounds with the long range magnetic ordering. In the LDA+U approach
the potential is defined \cite{b.anisimov,b.dudarev} as
\begin{equation}
V^{\rm LSDA+U}_{\sigma j l}=V^{\rm LSDA}+(U-J) [ \frac{1}{2}\delta_{jl}-\rho^{\sigma}_{jl}],
\end{equation}
where $\rho^{\sigma}_{jl}$ is the density matrix of $d$ electrons.

\begin{figure}
\resizebox*{7cm}{!}{\rotatebox{270}{\includegraphics{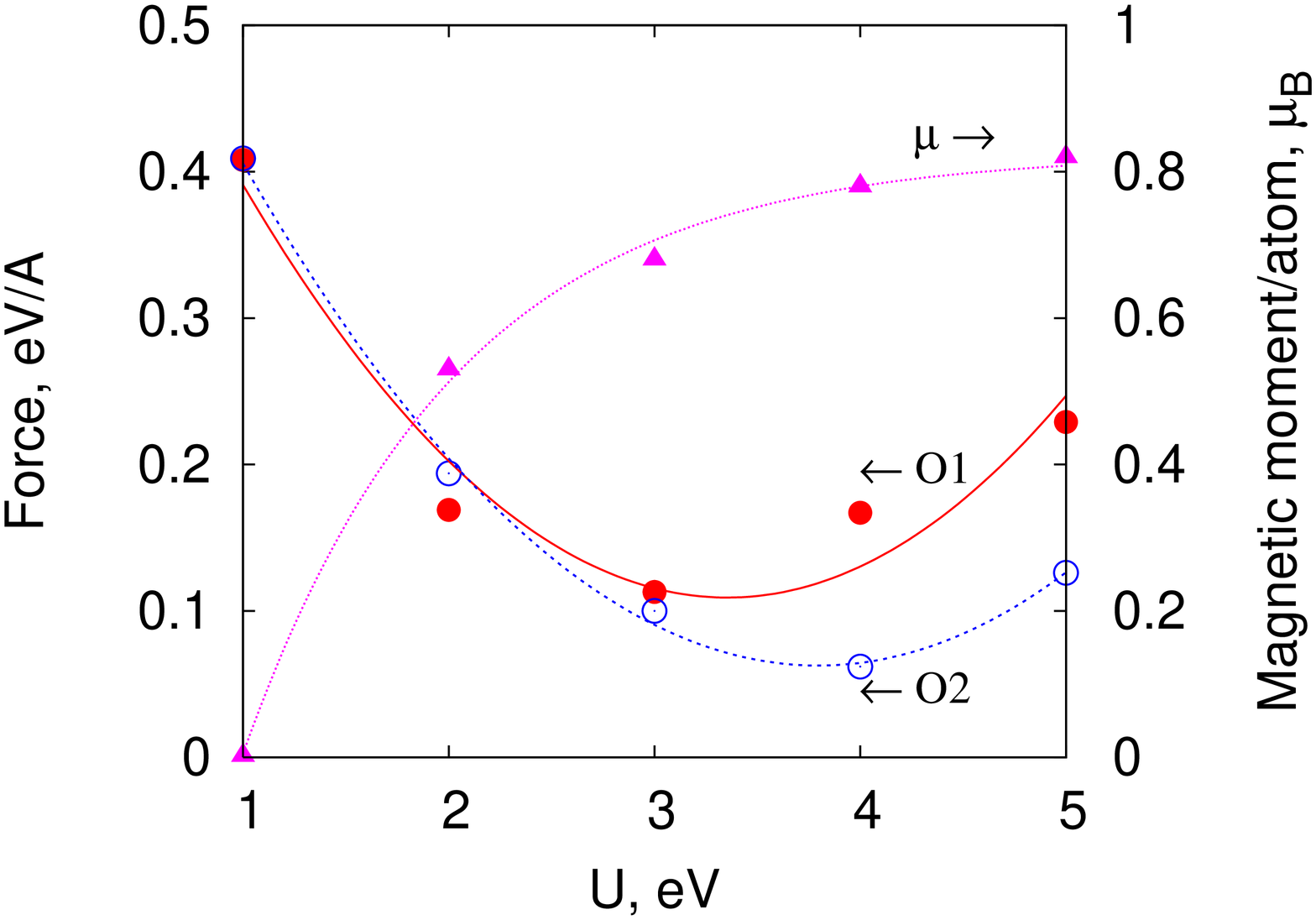}}}
\resizebox*{7cm}{!}{\rotatebox{270}{\includegraphics{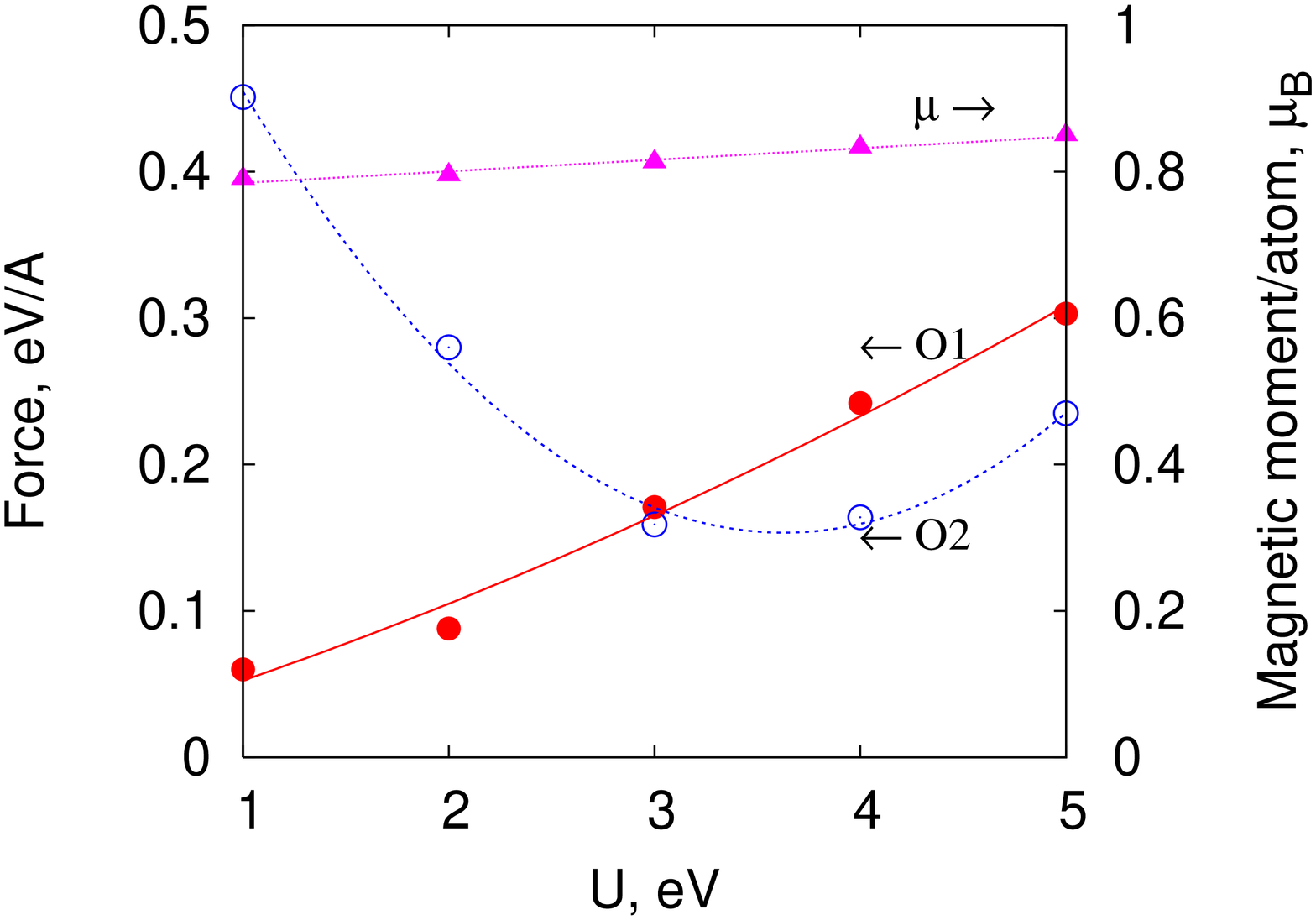}}}

\caption{({\it color online}) The one site electron-electron interaction
energy ($U$) dependency of the modules of the forces acting on different
atoms in LaTiO$_3$ ({\it left}) and YTiO$_3$ ({\it right}). }
\label{f.fu}
\end{figure}

The value of the Coulomb parameter $U^{\rm Ti}_{3d}$ was varied from 1
up to 5~eV. Since the value of the intra-atomic Hunds rule exchange
parameter is almost independent on materials \cite{b.anisimov} we use a
fixed value $J=1$~eV. The results of the GGA+U calculations for
LaTiO$_3$ and YTiO$_3$ with $U^{\rm Ti}_{3d}$=3~eV are shown by black color in
fig.~\ref{f.dos}. In both cases the oxygen bands are slightly higher in
energy $\sim 0.1$~eV in comparison with the GGA results. The $t_{2g}$
band is splitted and the value of the gap is equal to 0.4~eV and 0.6~eV
for LaTiO$_3$ and YTiO$_3$ respectively. The magnetic moment per Ti ion in LaTiO$_3$
grows fast with the value of Coulomb interaction and saturates to
$0.83\mu_{\rm B}$ at $U^{\rm Ti}_{3d}\simeq 4$~eV (see fig. \ref{f.fu}).
The experimental magnitude of the magnetic moment for LaTiO$_3$ corresponds
to $U^{\rm Ti}_{3d}\simeq 2$~eV in our calculations. The insulating gap
appears at $U^{\rm Ti}_{3d}\simeq 3$~eV. In contrast to the GGA
calculation the ground state of LaTiO$_3$ is the G-type AFM with the energy
gain relative to other types of magnetic ordering of 68 (FM), 85 (C-type
AFM) and 23~meV (A-type AFM). In YTiO$_3$ the value of gap increases with
the magnitude of the Coulomb interaction ({\it not shown}). The
experimental gap of 1.2~eV wide corresponds to parameter $U^{\rm
Ti}_{3d}= 4$. The magnetic moment in YTiO$_3$ (fig.~\ref{f.fu}) is almost
constant and equal to $\sim 0.8$~eV. The ferromagnetic solution is the
lowest in energy.

In fig.~\ref{f.fu} we plot the modules of the forces acting on O1 and O2
atoms versus the value of the Coulomb interaction. One can see the
forces for O1 and O2 in LaTiO$_3$ and O2 in YTiO$_3$ reduce with increasing of
$U^{\rm Ti}_{3d}$ and reach their minimum at about 3.5~eV. The forces
for O1 in YTiO$_3$ increase with $U^{\rm Ti}_{3d}$. This fact can be related
to the high temperature structure used, thus we expect large distortions
in YTiO$_3$ below $T_{\rm C}$. The value of $U^{\rm Ti}_{3d}=3-4$~eV is in
good agreement with the previous experimental \cite{b.okimoto,b.ulrich}
and theoretical \cite{b.solovyev1} estimations.

\begin{figure}
\begin{tabular}{cc}
LaTiO$_3$&YTiO$_3$\\
\resizebox*{!}{7cm}{\rotatebox{0}{\includegraphics{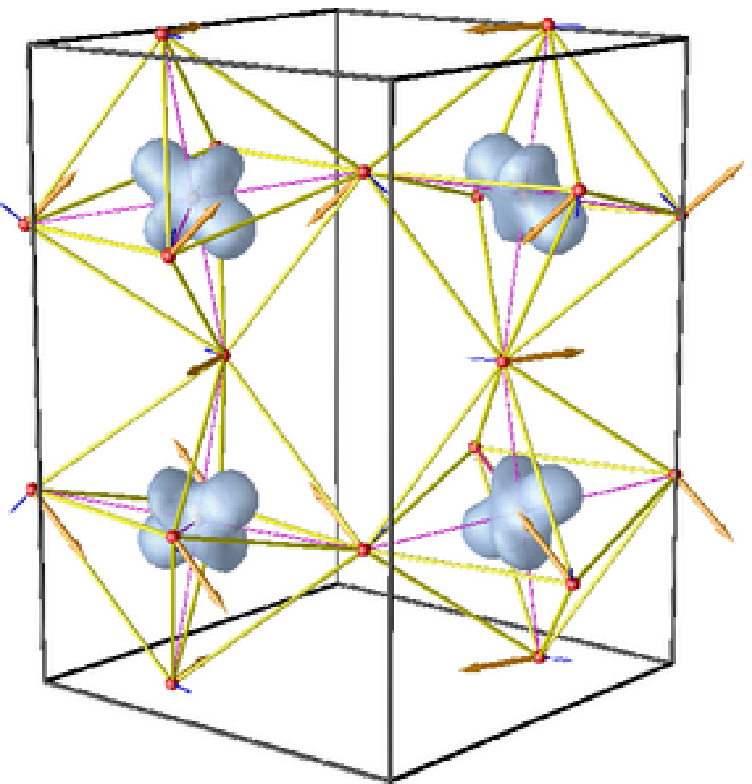}}}&
\resizebox*{!}{7cm}{\rotatebox{0}{\includegraphics{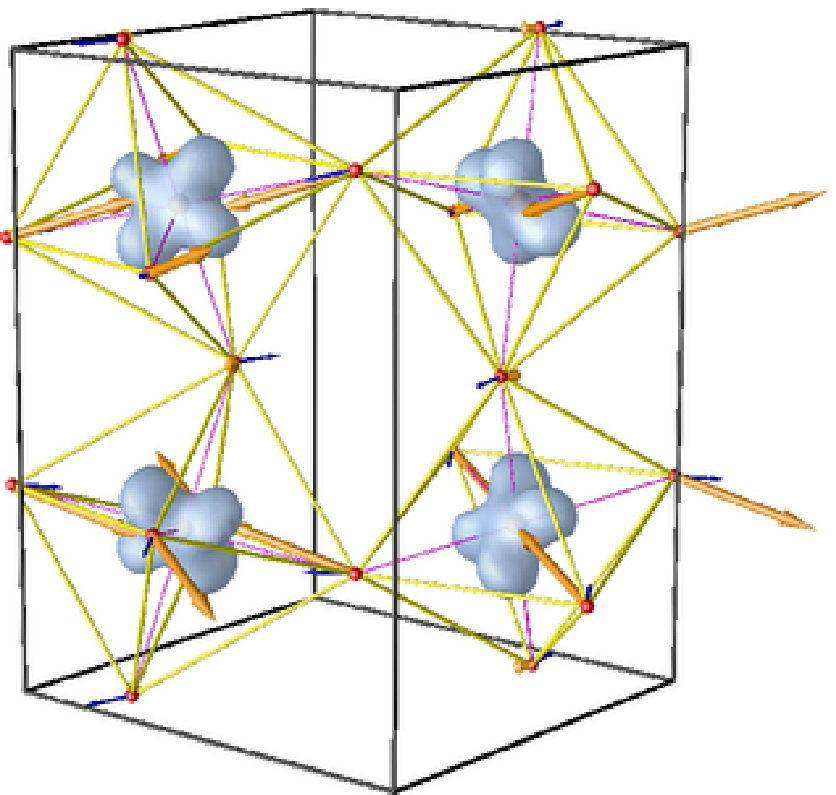}}}\\
\end{tabular}
\caption{({\it color online}) Charge densities in the energy window from
-2~eV till EF ({\rm gray surface}). Arrows are forces acting on atoms in
GGA ({\it orange}) and GGA+U with $U^{\rm Ti}_{3d}=3$ ({\it blue}).
$U^{\rm La}_{4f}=8$~eV. }
\label{f.dens}
\end{figure}

The directions of the forces calculated in the GGA and GGA+U ($U^{\rm
Ti}_{3d}=3$~eV) approximations are shown in fig.~\ref{f.dens} by yellow
and blue colors respectively. In the case of the GGA method one can see
that in LaTiO$_3$ they are almost equal in absolute value and trend more to
rotate the octahedra than to distort them. In YTiO$_3$ the two inward forces
face nearly along Ti--O bonds, while the two rest outward ones rotate.
Thus the GGA method tries to make the structure more cubic. That is a
characteristic of the titanates at temperatures above the orbital
ordering (700~K \cite{b.hemberger}) where the electrons become more
delocalized and are better described by the GGA. In the case of GGA+U
the forces smoothly decrease and rotate with increasing of $U^{\rm
Ti}_{3d}$ ({\it not shown}). At $U^{\rm Ti}_{3d}=3$~eV the angle between
GGA and GGA+U forces is $\sim 90^o$ in LaTiO$_3$ and  $\sim 30^o$ in YTiO$_3$
(fig.~\ref{f.dens}).

In order to visualize the resulting orbital ordering we plot the charge
density in fig.~\ref{f.dens} for both titanates integrated in the energy
window from $-2$~eV till Fermi level. In this energy interval the charge
density has predominant Ti-$t_{2g}$ character. One can see different
types of ordering in LaTiO$_3$ and YTiO$_3$: whereas in LaTiO$_3$ the orbitals arrange
in a ``fish bone'' style they form a rhombus in YTiO$_3$. These results
agree with those reported in previous theoretical findings
\cite{b.pavarini} and experimental works \cite{b.akimitsu, b.kiyama,
b.ichikawa, b.cwik}. The presented electronic densities clarify the
mechanism of the lattice distortions. Because of the repulsion between
negatively charged orbitals and oxygen atoms the different types of OO
cause differnt types of the distortions. The orbitals in LaTiO$_3$ do not
face toword any oxigens, thus distortions of the octahedra are small
there and their rotations are dominant. On contrary in YTiO$_3$ a specific
elongation of the orbitals toword O-atoms makes strong distortions of
the octahedra favourable. Since OO in the GGA calculations is weak the
corresponding forces face face opposite direction. Thus taking into
account the one site electronic correlations with the energy of Coulomb
interaction around $3-4$~eV we obtain a correct type of the orbital
ordering and can describe the crystal distortions.

\section{Conclusions}

We have found the correct magnetic and electronic ground state for LaTiO$_3$
and YTiO$_3$ in the framework of the GGA+U calculations. The value of
Coulomb interaction parameter estimated from the minimum of the forces
is about $3-4$~eV. This is in good agreement with the previous
theoretical studies and photoemission experiments \cite{b.mizokawa}. In
YTiO$_3$ in the GGA+U method with the optimal $U$-values the band-gap is
$0.6-1.2$~eV and the magnetic moment is $\mu = 0.81-0.83\mu_{\rm B}$ in
a good agreement with experimental data. The values of $\mu =
0.7\mu_{\rm B}$ and gap $E_{\rm g}=0.4$~eV for $U^{\rm Ti}_{3d}=3$~eV in
LaTiO$_3$ are overestimated because of the absence of the magnetic
fluctuations in GGA+U \cite{b.pavarini}. The orbital order found for all
values of $U^{\rm Ti}_{3d}$ is ``fish-bone'' and rhombus-type for LaTiO$_3$
and YTiO$_3$ respectively and is consistent with experimental structure
findings.

\acknowledgments

The authors acknowledge G.~de~Wijs, G.~Kresse for the technical help and
V.I.~Anisimov, S.~Streltsov for the useful discussions. We thank also
Stichting FOM (the Netherlands) for the financial support.


\end{document}